\documentclass[aps,twocolumn,showpacs,prc]{revtex4}
\usepackage{graphics,amssymb,epsfig}
\newcommand{\cbeta}{{\cal B}}
\begin{document}
\bibliographystyle{apsrev}
\title{Classification of the Nuclear Multifragmentation Phase Transition}
\author{Oliver M{\"u}lken}
\author{Peter Borrmann}
\affiliation{Department of Physics, Carl von Ossietzky University
Oldenburg, D-26111 Oldenburg, Germany}
\date{\today}
\begin{abstract}
Using a recently proposed classification scheme for phase transitions 
in finite systems [Phys.Rev.Lett.{\bf 84},3511 (2000)] 
we show that within the statistical standard model of nuclear
multifragmentation the predicted phase transition is of first order. 
\end{abstract}
\pacs{21.65.+f, 05.70.Fh, 25.70.Pq}
\maketitle
\section{Introduction} \label{intro}
The phenomenon of heavy-ion collision induced nuclear
multifragmentation (NMF) has been extensively studied over the past
decade.  Experimentally it is found that nuclei with excitation
energies of $E^*/A = $ 5-10 MeV, i.e.~in the region of the binding
energy, expand forming a number of intermediate mass fragments
\cite{Nebauer:1999}. Using subtle and advanced experimental
techniques  from a number of collision experiments with different
nuclei and at different collision energies an experimental caloric
curve has been derived\cite{Pochodzalla:1995,Ma:1997}, which
supports the interpretation that NMF is the small particle number
onset of the nuclear liquid-gas phase transition.  Theoretically NMF
has been described by a number of different approaches ranging from
simple percolation models\cite{Campi:1988,Bauer:1995}, dynamical
models with different levels of sophistication, e.g.~quantum
molecular dynamics of different flavors \cite{Aichelin:1986,
Hartnack:1989, Maruyama:1992, Bondorf:1995b,Peilert:1992} and
fermion molecular dynamics \cite{Feldmeier:2000}, and a large
variety of statistical models \cite{Bondorf:1995, Gross:1997,
Parvan:2000, Bhattacharyya:1999, Elliott:2000, Chase:1995, Lee:1992,
DeAngelis:1989}. Among the statistical models two
mainstream-theories, namely the {\it statistical multifragmentation
model} (SMM) \cite{Bondorf:1995} using the heat bath or canonical
ensemble and the {\it micro-canonical multifragmentation model} (MMM)
\cite{Gross:1997} using the constant energy or micro-canonical
ensemble should be mentioned. For different ensembles the shape of
the caloric curve may differ significantly. However, because all ensembles
are uniquely connected by simple integral transforms the qualitative
features of the phase transition should be the same in all ensembles. We
therefore omit at this point the (surely) interesting question
which ensemble is more appropriate for the description of nuclear
multifragmentation. 

In this paper we use the canonical statistical multifragmentation
model based on the works of Mekjian {\sl et al.}
\cite{DeAngelis:1989,Mekjian:1990,Lee:1992,Lee:1993,Mekjian:1991a,
Chase:1994,Chase:1995} to analyze the order of the
multifragmentation phase transition using a recently proposed
classification scheme for phase transitions in finite
systems~\cite{Borr99b,Muelken:2000}. The classification scheme is
based on an analysis of the distribution of zeros of the canonical
partition function in the complex plane and is in the thermodynamic
limit in complete accordance with the Ehrenfest classification
scheme. This scheme is shortly described in Sec.~\ref{class}. In
Sec.~\ref{model} we briefly review the statistical model used in
this paper. The results of our
analysis which clearly show that NMF has the signature of a first
order phase transition are presented in Sec.~\ref{results}. 
%
%
\section{classification scheme}\label{class}
The classification scheme used below and described in far more
detail in \cite{Borr99b,Muelken:2000} is based on the analytic
continuation of the inverse temperature $\beta$ into the complex
plane $\beta\to\cbeta=\beta+i\tau$. For macroscopic systems an
equivalent scheme was developed by Grossmann {\sl et
al.}~\cite{Gross1967}. All
thermodynamic information about a system can be extracted from the
distribution of complex zeros of the canonical partition function
$Z(\cbeta) = \int {\rm d}E \ \Omega(E) \ \exp(-\cbeta E)$.  

To avoid some difficulties with the high temperature limit we write
the partition function as a product $Z(\cbeta)=Z_{\rm lim}(\cbeta)
Z_{\rm int}(\cbeta)$, explicitly  separating  the high temperature
limit ($T\to\infty$) of the partition function $Z_{\rm lim}(\cbeta)$
imposing $\lim_{T\to\infty} Z_{\rm int}(\cbeta)=1$. 
The partition function is an integral function with complex
conjugate zeros $\cbeta_k = \cbeta_{-k}^* = \beta_k + i \tau_k \ (k
\in \mathbb{N})$ and can be rewritten as 
\begin{eqnarray} 
Z(\cbeta) &=& Z_{\rm lim}(\cbeta) \ Z_{\rm int}(0) \ \exp(\cbeta
\partial_\cbeta \ln Z_{\rm int}(0)) \\ &\times&
\prod_{k\in\mathbb{N}} \left( 1-\frac{\cbeta}{\cbeta_k} \right)
\left( 1-\frac{\cbeta}{\cbeta_k^*} \right) \ \exp\left(
\frac{\cbeta}{\cbeta_k} + \frac{\cbeta}{\cbeta_k^*} \right).
\nonumber 
\end{eqnarray}
All thermodynamic properties like the internal energy or specific heat are
derivatives of the logarithm of the canonical partition function. Thus, the
zeros of $Z(\cbeta)$ are poles of these quantities. In the thermodynamic limit
different regions of holomorphy separate different phases by dense lines of
zeros.  In finite systems the zeros do not become dense on lines which leads to
a less sharp separation of different phases. We interpret the zeros as
boundary posts between two phases.  
\begin{figure}
\centerline{\epsfig{file=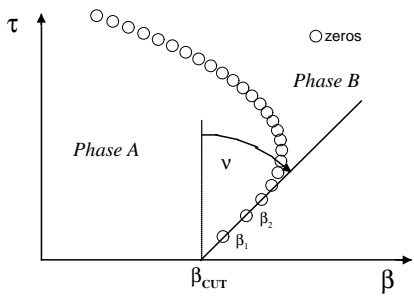,width=4cm,angle=270,clip=}}
\caption{Schematic illustration of the zeros in the complex
temperature plane.}
\label{schema}
\end{figure}

The distribution of zeros close to the real
axis can approximately be described by three parameters, where two of them
reflect the order of the phase transition and the third merely the size of the
system. We assume that the zeros lie on straight lines with a discrete density
of zeros given by
\begin{equation}
\phi(\tau_k)=\frac12\left(\frac{1}{|\cbeta_k-\cbeta_{k-1}|}+
\frac{1}{|\cbeta_{k+1}-\cbeta_k|}\right),\label{discretedensity}
\end{equation}
with $k=2,3,4,\cdots$, and approximate for small $\tau$ the density
of zeros by a simple power law $\phi(\tau)\sim\tau^\alpha$.
Considering only the first three zeros the exponent $\alpha$ can be
estimated as
\begin{equation}
\alpha=\frac{\ln\phi(\tau_3)-\ln\phi(\tau_2)}{\ln\tau_3-\ln\tau_2}.
\end{equation}
The second parameter describing the distribution of zeros is given
by $\gamma=\tan\nu \sim (\beta_2-\beta_1)/ (\tau_2-\tau_1)$ where
$\nu$ is the crossing angle of the line of zeros with the real axis
(see Fig.~\ref{schema}).  The {\sl discreteness} of the system is
reflected in the imaginary part $\tau_1$ of the zero closest to the
real axis.
For macroscopic systems we always have $\tau_1\to0$. In this case
the parameters $\alpha$ and $\gamma$ match those defined by
Grossmann {\sl et al.}~\cite{Gross1967}, who showed the connection
between certain values of $\alpha$ and $\gamma$ and different types
of phase transitions. They claimed that a first order transition
occurs for $\alpha=\gamma=0$, for $0<\alpha<1$ with $\gamma=0$ or
$\gamma\neq0$ a second order transition, and for $\alpha>1$ a higher
order transition. In the thermodynamic limit $\alpha$ is restricted
to values greater than or equal to zero because otherwise this would cause a
diverging internal energy. For finite systems values of $\alpha$
less than zero are possible.

In our classification scheme phase transitions in finite systems are
of first order if $\alpha\leq0$, while the definitions for higher 
order transitions coincide with those given by Grossmann. 
Additionally we define $\tau_1$ as an unambiguous parameter for the
discreteness of the system. Since in small systems no thermodynamic 
properties diverge, the specification of the critical
temperature is difficult and several definitions are possible which
coincide only in the thermodynamic limit. We define the real part of the
first zero $\beta_1$ as the critical temperature. Another possible 
definition would be the crossing point of the line of zeros with 
the real temperature axis.
%
%
\section{statistical multifragmentation model}\label{model}
In the statistical models for NMF~\cite{Bondorf:1995, Chase:1994,
Chase:1995, Mekjian:1991a} it is assumed that the fragmentation process
can be described in the following manner. After initial excitation the
system expands and fragments into a number of clusters.  During the
expansion a thermalization of all clusters is assumed.  The thermalization
abruptly stops when the mean separation between the clusters exceeds the
range of the nuclear force. The properties of the systems at this
so-called {\sl freeze-out} density, which is assumed to be 0.1-0.3 of the
normal nuclear density  $\rho_0=0.15$ fm$^{-3}$, are than carried by all
fragments until experimental detection.

The probability for a fragmentation with fragmentation
vector $\vec n = (n_1,n_2,\dots)$ ($n_k$ is the number of
clusters of size $k$ and $\sum k n_k = A$), reads
\begin{equation}
W(\vec n) = \prod_{k\geq 1} \frac{1}{n_k!} 
\left( \frac{x y^{k-1}}{\alpha_k} \right)^{n_k},
\end{equation}
where $x=V_{\rm ac}\left(\frac{M} {2\pi\hbar^2\cbeta}\right)^{3/2}$
is the partition function for the ideal monoatomic gas
of mass $M$ in the accessible volume $V_{\rm ac}$. 
$y=\exp\big(a_v(\cbeta)\cbeta+\frac{T_0/\epsilon_0}{1+T_0\cbeta}\big)$
is related to the cluster binding and internal excitations. The cluster
size dependence of the probability is given by $\alpha_k = k^{5/2}$.
We assume an accessible volume three times larger than the normal
nucleus volume $V_0$. The canonical partition function is then given 
as 
\begin{figure*}[ht]
\centerline{\epsfig{file=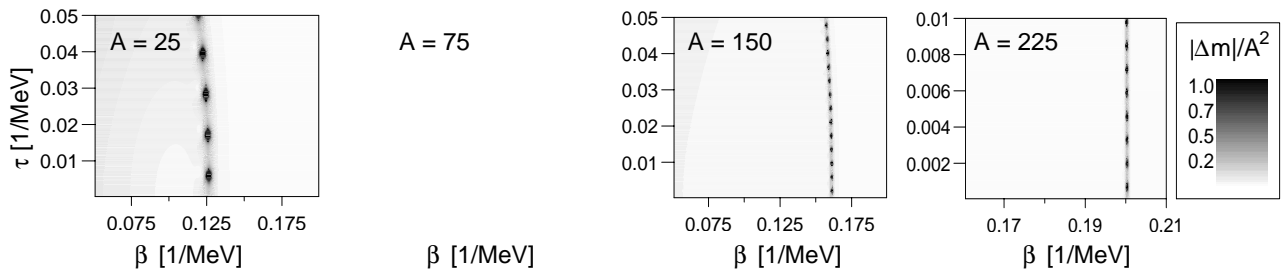,width=17cm,clip=}}
\caption{Contour plots of the variance of the multiplicity in the
complex temperature plane for (a) A=25, (b) A=75, (c) A=150, and (d) A=225
nucleons. The dark spots indicate the locations of the zeros of
$Z(\cbeta)$.}\label{multizeros}
\end{figure*}
\begin{equation}
Z_A = \sum_{m=0}^A Z_A^{(m)} x^m y^{A-m},
\end{equation}
where $m = \sum_k n_k$ is the multiplicity, i.e.~the total number of
clusters, and $Z_A^{(m)}$ is independent of the thermodynamic
variables and given by the simple recursion
\begin{equation} 
Z_A^{(m)}=\frac1m \sum_{k=1}^A \frac{1}{\alpha_k} Z_{A-k}^{(m-1)},
\label{recursion}
\end{equation}
with $Z_A^{(1)} = 1/\alpha_A$. 

Following Ref.~\cite{Bondorf:1995} the binding energy for a
composite of $k$ nucleons is given by
\begin{equation}
a_v(\cbeta) = W_0 k - \sigma(\cbeta) k^{2/3},
\label{binding}
\end{equation}
with $\sigma(\cbeta) = \sigma_0\left((T_c-1/\cbeta^2)
/(T_c+1/\cbeta^2) \right)^{5/4}$ being the surface tension ($W_0 =
16$MeV, $\sigma_0=18$MeV, and $T_c=18$MeV).

Within this framework different moments of the multiplicity are
calculated by
\begin{equation}
\left<m^i\right> = 
\frac{1}{Z_A} \sum_{m=0}^A m^i Z_A^{(m)} x^m y^{A-m}.
\end{equation}
The variance of the multiplicity is given by $\Delta m=
\langle m^2\rangle-\langle m\rangle^2$. It is straightforward to
express thermodynamic quantities like the internal energy or the
specific heat in terms of different moments of the
multiplicity~\cite{Chase:1995}.
%
%
\section{results}\label{results}
Due to the unfavorable scaling behavior of the partition
function it is very arduous to find its zeros directly with
numerical methods. The zeros of $Z$ are the poles of almost all
thermodynamic expectation values, e.g. the internal energy, the
specific heat and so on. In this paper we utilize the absolute value
of the variance of
the multiplicity $|\langle \Delta m \rangle|$
to locate the zeros of $Z$
in the complex temperature plane. Fig.~\ref{multizeros} shows a
contour plot of $|\langle\Delta m\rangle|/A^2$ in the complex temperature
plane for nucleon-numbers $A=25,75,150$, and $225$ (note the
different scaling of the $\tau$-axis.). This figure directly 
shows that multifragmentation is a first order phase transition.
The zeros lie on straight lines perpendicular to the real temperature 
axis with constant spacings between them. Even at particle numbers as 
small as 25 the nature of the phase transition can be clearly
identified from the lines of zeros of the partition function.
\begin{figure}[b]
\centerline{\epsfig{file=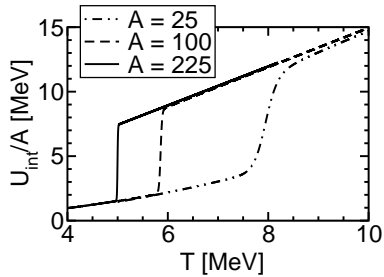,width=5cm,clip=}}
\caption{Canonical caloric curve for A=25, 100, and A=225.}\label{caloric}
\end{figure}
In contrast, it is very difficult to extract the order of the phase 
transition from the canonical caloric curve (internal energy U$_{\rm int}$
per particle number A vs.~temperature T) shown in Fig.~\ref{caloric}.

The recursion formula (\ref{recursion}) has a very similar structure
to that for Bose-Einstein systems \cite{bf93,Borr99a,Muelken:2000}.
This is because both systems have an intimate relationship to the
permutation group \cite{Mekjian:1991a}. It is therefore remarkable,
that the multifragmentation phase transition for nuclear matter is
of first order while Bose-Einstein condensation is a second or third
order phase transition depending on the single particle density of
states.

\begin{figure}
\centerline{\epsfig{file=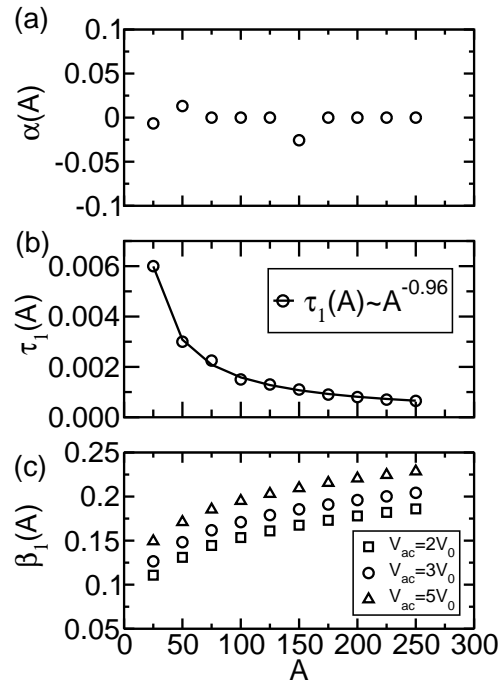,width=6.5cm,clip=}}
\caption{Nucleon number dependent value of the classification parameters
(a) $\alpha$, (b) $\tau$ for an accessible volume $V_{\rm ac}=3 V_0$ and
(c) of the critical temperature $\beta_1$ for three different accessible
volumes.}
\label{param}
\end{figure}

A detailed analysis of the distribution of zeros reveals that for all
nucleon numbers ($A>25$) the most important classification parameter
$\alpha$ is almost equal to zero (see Fig.~\ref{param}(a)). The slight
deviations for A=50 and A=150 are most likely due to uncertainties within
our numerical zero detection method. The second parameter $\gamma$ can
directly be extracted from Fig.~\ref{multizeros}. Even for small nucleon
numbers the distribution of zeros is perpendicular to the real axis,
corresponding to $\gamma=0$.  The finite size of the system is connected
to the third classification parameter $\tau_1$.  Fig.~\ref{param}(b) shows
the $A$-dependence of $\tau_1$.  With increasing nucleon number $\tau_1$
decreases. We approximated the scaling behavior as $\tau_1 \sim
A^{-0.96}$, i.e.~$\tau_1$ is simply volume dependent and the phase
transition would approach a true first order phase transition in the
Ehrenfest sense with $1/A$.  The critical temperature $T=1/\beta_1$
(Fig.~\ref{param}(c)) decreases and adopts at $A\approx200$ the
experimentally found value of about $5$ MeV.  Of course, for larger system
sizes the simple equation (\ref{binding}) for the binding energy used in
this study should be refined and augmented by additional terms. Additional
calculations for different particle densities reveal that there are no
qualitative changes to the first order nature of the phase transition.
Fig.~\ref{param}(c) also shows calculations for accessible volumes which
are twice, three, and five times as large as the normal nucleus volume
$V_0$. There are only quantitative changes in the critical temperature.

It should be noted that one can expect that refinements of the
model, e.g.~the introduction of an additional Coulomb term and other
values of the particle density, may influence the details of the
distribution of zeros. Especially tuning the particle density to
the expected supercritical phase is an interesting challenge, worth to be
pursued in the future.
%
%
\section{conclusion}
In conclusion we have found that within the statistical model nuclear
multifragmentation is undoubtedly a first order phase transition, which is
in complete agreement with recent works based on micro-canonical and
grand-canonical calculations~\cite{Chomaz:2000,Bugaev:2000}. Of course,
the first order type of the NMF phase transition is not yet sufficient to
interprete NMF as the onset of the nuclear liquid-gas transition. However,
this seems to be a necessary condition. The distribution of zeros of the
partition function in the complex temperature plane proofs as a reliable
tool for the classification of phase transitions, adding a clear,
detailed, and unambiguous view on the thermodynamic properties of small
systems. 

\begin{acknowledgments}
We wish to thank Jens Harting, Heinrich Stamerjohanns and
E.~R.~Hilf for fruitful discussions.
\end{acknowledgments}
%
%


\begin{thebibliography}{10}
\expandafter\ifx\csname bibnamefont\endcsname\relax
  \def\bibnamefont#1{#1}\fi
\expandafter\ifx\csname bibfnamefont\endcsname\relax
  \def\bibfnamefont#1{#1}\fi
\expandafter\ifx\csname url\endcsname\relax
  \def\url#1{\texttt{#1}}\fi
\expandafter\ifx\csname urlprefix\endcsname\relax\def\urlprefix{URL }\fi
\expandafter\ifx\csname bibinfo\endcsname\relax \def\bibinfo#1#2{#2}\fi
\expandafter\ifx\csname eprint\endcsname\relax \def\eprint#1{#1}\fi

\bibitem{Nebauer:1999}
\bibinfo{author}{\bibfnamefont{R.}~\bibnamefont{Nenauer}},
  \bibinfo{author}{\bibfnamefont{A.}~\bibnamefont{J.}}, \bibnamefont{and}
  \bibinfo{author}{\bibnamefont{the INDRA~collaboration}},
  \bibinfo{journal}{Nucl. Phys. A} \textbf{\bibinfo{volume}{658}},
  \bibinfo{pages}{67} (\bibinfo{year}{1999}), \bibinfo{note}{and references
  therein}.

\bibitem{Pochodzalla:1995}
\bibinfo{author}{\bibfnamefont{J.}~\bibnamefont{Pochodzalla}},
  \bibinfo{author}{\bibfnamefont{T.}~\bibnamefont{M{\"o}llenkamp}},
  \bibinfo{author}{\bibfnamefont{T.}~\bibnamefont{Rubehn}},
  \bibinfo{author}{\bibfnamefont{A.}~\bibnamefont{Sch{\"u}ttauf}},
  \bibinfo{author}{\bibfnamefont{W.}~\bibnamefont{A.}},
  \bibinfo{author}{\bibfnamefont{E.}~\bibnamefont{Zude}},
  \bibinfo{author}{\bibfnamefont{M.}~\bibnamefont{Begemann-Blaich}},
  \bibinfo{author}{\bibfnamefont{T.}~\bibnamefont{Blaich}},
  \bibinfo{author}{\bibfnamefont{H.}~\bibnamefont{Emling}},
  \bibinfo{author}{\bibfnamefont{A.}~\bibnamefont{Ferrero}},
  \bibinfo{author}{\bibfnamefont{C.}~\bibnamefont{Gross}},
  \bibinfo{author}{\bibfnamefont{G.}~\bibnamefont{Imme}}, \emph{et~al.},
  \bibinfo{journal}{Phys. Rev. Lett.} \textbf{\bibinfo{volume}{75}},
  \bibinfo{pages}{1040} (\bibinfo{year}{1995}).

\bibitem{Ma:1997}
\bibinfo{author}{\bibfnamefont{Y.-G.} \bibnamefont{Ma}},
  \bibinfo{author}{\bibfnamefont{A.}~\bibnamefont{Siwek}},
  \bibinfo{author}{\bibfnamefont{J.}~\bibnamefont{Peter}},
  \bibinfo{author}{\bibfnamefont{F.}~\bibnamefont{Gulminelli}},
  \bibinfo{author}{\bibfnamefont{R.}~\bibnamefont{Dayras}},
  \bibinfo{author}{\bibfnamefont{L.}~\bibnamefont{Nalpas}},
  \bibinfo{author}{\bibfnamefont{B.}~\bibnamefont{Tamain}},
  \bibinfo{author}{\bibfnamefont{E.}~\bibnamefont{Vient}},
  \bibinfo{author}{\bibfnamefont{G.}~\bibnamefont{Auger}},
  \bibinfo{author}{\bibfnamefont{C.}~\bibnamefont{Bacri}},
  \bibinfo{author}{\bibfnamefont{J.}~\bibnamefont{Benlliure}},
  \bibinfo{author}{\bibfnamefont{E.}~\bibnamefont{Bisquer}}, \emph{et~al.},
  \bibinfo{journal}{Phys. Lett. B} \textbf{\bibinfo{volume}{390}},
  \bibinfo{pages}{41} (\bibinfo{year}{1997}).

\bibitem{Campi:1988}
\bibinfo{author}{\bibfnamefont{X.}~\bibnamefont{Campi}},
  \bibinfo{journal}{Phys. Lett. B} \textbf{\bibinfo{volume}{208}},
  \bibinfo{pages}{351} (\bibinfo{year}{1988}).

\bibitem{Bauer:1995}
\bibinfo{author}{\bibfnamefont{W.}~\bibnamefont{Bauer}} \bibnamefont{and}
  \bibinfo{author}{\bibfnamefont{A.}~\bibnamefont{Botvina}},
  \bibinfo{journal}{Phys. Rev. C} \textbf{\bibinfo{volume}{52}},
  \bibinfo{pages}{1750} (\bibinfo{year}{1995}).

\bibitem{Aichelin:1986}
\bibinfo{author}{\bibfnamefont{J.}~\bibnamefont{Aichelin}} \bibnamefont{and}
  \bibinfo{author}{\bibfnamefont{H.}~\bibnamefont{St{\"o}cker}},
  \bibinfo{journal}{Phys. Lett. B} \textbf{\bibinfo{volume}{176}},
  \bibinfo{pages}{14} (\bibinfo{year}{1986}).

\bibitem{Hartnack:1989}
\bibinfo{author}{\bibfnamefont{C.}~\bibnamefont{Hartnack}},
  \bibinfo{author}{\bibfnamefont{L.}~\bibnamefont{Zhuxia}},
  \bibinfo{author}{\bibfnamefont{L.}~\bibnamefont{Neise}},
  \bibinfo{author}{\bibfnamefont{G.}~\bibnamefont{Peilert}},
  \bibinfo{author}{\bibfnamefont{A.}~\bibnamefont{Rosenhauer}},
  \bibinfo{author}{\bibfnamefont{H.}~\bibnamefont{Sorge}},
  \bibinfo{author}{\bibfnamefont{J.}~\bibnamefont{Aichelin}},
  \bibinfo{author}{\bibfnamefont{H.}~\bibnamefont{St{\"o}cker}},
  \bibnamefont{and} \bibinfo{author}{\bibfnamefont{W.}~\bibnamefont{Greiner}},
  \bibinfo{journal}{Nucl. Phys. A} \textbf{\bibinfo{volume}{495}},
  \bibinfo{pages}{303} (\bibinfo{year}{1989}).

\bibitem{Maruyama:1992}
\bibinfo{author}{\bibfnamefont{T.}~\bibnamefont{Maruyama}},
  \bibinfo{author}{\bibfnamefont{K.}~\bibnamefont{Niita}}, \bibnamefont{and}
  \bibinfo{author}{\bibfnamefont{A.}~\bibnamefont{Iwamoto}},
  \bibinfo{journal}{Phys. Rev. C} \textbf{\bibinfo{volume}{53}},
  \bibinfo{pages}{297} (\bibinfo{year}{1996}).

\bibitem{Bondorf:1995b}
\bibinfo{author}{\bibfnamefont{J.}~\bibnamefont{Bondorf}},
  \bibinfo{author}{\bibfnamefont{D.}~\bibnamefont{Idier}}, \bibnamefont{and}
  \bibinfo{author}{\bibfnamefont{I.}~\bibnamefont{Mishustin}},
  \bibinfo{journal}{Phys. Lett. B} \textbf{\bibinfo{volume}{359}},
  \bibinfo{pages}{261} (\bibinfo{year}{1995}).

\bibitem{Peilert:1992}
\bibinfo{author}{\bibfnamefont{G.}~\bibnamefont{Peilert}},
  \bibinfo{author}{\bibfnamefont{J.}~\bibnamefont{Konopka}},
  \bibinfo{author}{\bibfnamefont{M.}~\bibnamefont{Mustafa}},
  \bibinfo{author}{\bibfnamefont{H.}~\bibnamefont{St{\"o}cker}},
  \bibnamefont{and} \bibinfo{author}{\bibfnamefont{W.}~\bibnamefont{Greiner}},
  \bibinfo{journal}{Phys. Rev. C} \textbf{\bibinfo{volume}{46}},
  \bibinfo{pages}{1457} (\bibinfo{year}{1992}).

\bibitem{Feldmeier:2000}
\bibinfo{author}{\bibfnamefont{H.}~\bibnamefont{Feldmeier}} \bibnamefont{and}
  \bibinfo{author}{\bibfnamefont{J.}~\bibnamefont{Schnack}},
  \bibinfo{journal}{Rev. Mod. Phys.} \textbf{\bibinfo{volume}{72}},
  \bibinfo{pages}{655} (\bibinfo{year}{2000}).

\bibitem{Bondorf:1995}
\bibinfo{author}{\bibfnamefont{J.}~\bibnamefont{Bondorf}},
  \bibinfo{author}{\bibfnamefont{B.}~\bibnamefont{A.S.}},
  \bibinfo{author}{\bibfnamefont{A.}~\bibnamefont{Iljinov}},
  \bibinfo{author}{\bibfnamefont{I.}~\bibnamefont{Mishustin}},
  \bibnamefont{and} \bibinfo{author}{\bibfnamefont{K.}~\bibnamefont{Sneppen}},
  \bibinfo{journal}{Phys. Rep.} \textbf{\bibinfo{volume}{257}},
  \bibinfo{pages}{133} (\bibinfo{year}{1995}).

\bibitem{Gross:1997}
\bibinfo{author}{\bibfnamefont{D.}~\bibnamefont{Gross}},
  \bibinfo{journal}{Phys. Rep.} \textbf{\bibinfo{volume}{279}},
  \bibinfo{pages}{119} (\bibinfo{year}{1995}).

\bibitem{Parvan:2000}
\bibinfo{author}{\bibfnamefont{A.}~\bibnamefont{Parvan}},
  \bibinfo{author}{\bibfnamefont{V.}~\bibnamefont{Toneev}}, \bibnamefont{and}
  \bibinfo{author}{\bibfnamefont{M.}~\bibnamefont{Ploszajczak}},
  \bibinfo{journal}{Nucl. Phys. {\bf A}} \textbf{\bibinfo{volume}{676}},
  \bibinfo{pages}{409} (\bibinfo{year}{2000}).

\bibitem{Bhattacharyya:1999}
\bibinfo{author}{\bibfnamefont{P.}~\bibnamefont{Bhattacharyya}},
  \bibinfo{author}{\bibfnamefont{S.}~\bibnamefont{Das~Gupta}},
  \bibnamefont{and} \bibinfo{author}{\bibfnamefont{A.}~\bibnamefont{Mekjian}},
  \bibinfo{journal}{Phys. Rev. {\bf C}} \textbf{\bibinfo{volume}{60}},
  \bibinfo{pages}{064625} (\bibinfo{year}{1999}).

\bibitem{Elliott:2000}
\bibinfo{author}{\bibfnamefont{J.}~\bibnamefont{Elliott}} \bibnamefont{and}
  \bibinfo{author}{\bibfnamefont{A.}~\bibnamefont{Hirsch}},
  \bibinfo{journal}{Phys. Rev. {\bf C}} \textbf{\bibinfo{volume}{61}},
  \bibinfo{pages}{054605} (\bibinfo{year}{2000}).

\bibitem{Chase:1995}
\bibinfo{author}{\bibfnamefont{K.}~\bibnamefont{Chase}} \bibnamefont{and}
  \bibinfo{author}{\bibfnamefont{A.}~\bibnamefont{Mekjian}},
  \bibinfo{journal}{Phys. Rev. Lett.} \textbf{\bibinfo{volume}{75}},
  \bibinfo{pages}{4732} (\bibinfo{year}{1995}).

\bibitem{Lee:1992}
\bibinfo{author}{\bibfnamefont{S.}~\bibnamefont{Lee}} \bibnamefont{and}
  \bibinfo{author}{\bibfnamefont{A.}~\bibnamefont{Mekjian}},
  \bibinfo{journal}{Phys. Rev. {\bf C}} \textbf{\bibinfo{volume}{44}},
  \bibinfo{pages}{1284} (\bibinfo{year}{1992}).

\bibitem{DeAngelis:1989}
\bibinfo{author}{\bibfnamefont{A.}~\bibnamefont{DeAngelis}} \bibnamefont{and}
  \bibinfo{author}{\bibfnamefont{A.}~\bibnamefont{Mekjian}},
  \bibinfo{journal}{Phys. Rev. {\bf C}} \textbf{\bibinfo{volume}{40}},
  \bibinfo{pages}{105} (\bibinfo{year}{1989}).

\bibitem{Mekjian:1990}
\bibinfo{author}{\bibfnamefont{A.}~\bibnamefont{Mekjian}},
  \bibinfo{journal}{Phys. Rev. {\bf C}} \textbf{\bibinfo{volume}{41}},
  \bibinfo{pages}{2103} (\bibinfo{year}{1990}).

\bibitem{Lee:1993}
\bibinfo{author}{\bibfnamefont{S.}~\bibnamefont{Lee}} \bibnamefont{and}
  \bibinfo{author}{\bibfnamefont{A.}~\bibnamefont{Mekjian}},
  \bibinfo{journal}{Phys. Rev. {\bf C}} \textbf{\bibinfo{volume}{47}},
  \bibinfo{pages}{2266} (\bibinfo{year}{1993}).

\bibitem{Mekjian:1991a}
\bibinfo{author}{\bibfnamefont{A.}~\bibnamefont{Mekjian}} \bibnamefont{and}
  \bibinfo{author}{\bibfnamefont{S.}~\bibnamefont{Lee}},
  \bibinfo{journal}{Phys. Rev. {\bf A}} \textbf{\bibinfo{volume}{44}},
  \bibinfo{pages}{6294} (\bibinfo{year}{1991}).

\bibitem{Chase:1994}
\bibinfo{author}{\bibfnamefont{K.}~\bibnamefont{Chase}} \bibnamefont{and}
  \bibinfo{author}{\bibfnamefont{A.}~\bibnamefont{Mekjian}},
  \bibinfo{journal}{Phys. Rev. {\bf C}} \textbf{\bibinfo{volume}{49}},
  \bibinfo{pages}{2164} (\bibinfo{year}{1994}).

\bibitem{Borr99b}
\bibinfo{author}{\bibfnamefont{P.}~\bibnamefont{Borrmann}},
  \bibinfo{author}{\bibfnamefont{O.}~\bibnamefont{M{\"u}lken}},
  \bibnamefont{and} \bibinfo{author}{\bibfnamefont{J.}~\bibnamefont{Harting}},
  \bibinfo{journal}{Phys.\ Rev.\ Lett.} \textbf{\bibinfo{volume}{84}},
  \bibinfo{pages}{3511} (\bibinfo{year}{2000}).

\bibitem{Muelken:2000}
\bibinfo{author}{\bibfnamefont{O.}~\bibnamefont{M{\"u}lken}},
  \bibinfo{author}{\bibfnamefont{P.}~\bibnamefont{Borrmann}},
  \bibinfo{author}{\bibfnamefont{J.}~\bibnamefont{Harting}}, \bibnamefont{and}
  \bibinfo{author}{\bibfnamefont{H.}~\bibnamefont{Stamerjohanns}},
  \bibinfo{journal}{arXiV:cond-mat/0006293}  (\bibinfo{year}{2000}).

\bibitem{Gross1967}
\bibinfo{author}{\bibfnamefont{S.}~\bibnamefont{Grossmann}} \bibnamefont{and}
  \bibinfo{author}{\bibfnamefont{W.}~\bibnamefont{Rosenhauer}},
  \bibinfo{journal}{Z. Phys.} \textbf{\bibinfo{volume}{207}},
  \bibinfo{pages}{138} (\bibinfo{year}{1967});
\bibinfo{author}{\bibfnamefont{W.}~\bibnamefont{Bestgen}},
  \bibinfo{author}{\bibfnamefont{S.}~\bibnamefont{Grossmann}},
  \bibnamefont{and}
  \bibinfo{author}{\bibfnamefont{W.}~\bibnamefont{Rosenhauer}},
  \bibinfo{journal}{J. Phys. Soc. Jap.} \textbf{\bibinfo{volume}{26}},
  \bibinfo{pages}{115} (\bibinfo{year}{1969});
\bibinfo{author}{\bibfnamefont{S.}~\bibnamefont{Grossmann}},
  \bibinfo{journal}{Phys. Lett.}
  \textbf{\bibinfo{volume}{28A}}(\bibinfo{number}{2}), \bibinfo{pages}{162}
  (\bibinfo{year}{1968});
\bibinfo{author}{\bibfnamefont{S.}~\bibnamefont{Grossmann}} \bibnamefont{and}
  \bibinfo{author}{\bibfnamefont{V.}~\bibnamefont{Lehmann}},
  \bibinfo{journal}{Z. Phys.} \textbf{\bibinfo{volume}{218}},
  \bibinfo{pages}{449} (\bibinfo{year}{1969}).

\bibitem{bf93}
\bibinfo{author}{\bibfnamefont{P.}~\bibnamefont{Borrmann}} \bibnamefont{and}
  \bibinfo{author}{\bibfnamefont{G.}~\bibnamefont{Franke}},
  \bibinfo{journal}{J. Chem. Phys} \textbf{\bibinfo{volume}{98}},
  \bibinfo{pages}{2484} (\bibinfo{year}{1993}).

\bibitem{Borr99a}
\bibinfo{author}{\bibfnamefont{P.}~\bibnamefont{Borrmann}},
  \bibinfo{author}{\bibfnamefont{J.}~\bibnamefont{Harting}},
  \bibinfo{author}{\bibfnamefont{O.}~\bibnamefont{M{\"u}lken}},
  \bibnamefont{and} \bibinfo{author}{\bibfnamefont{E.}~\bibnamefont{Hilf}},
  \bibinfo{journal}{Phys. Rev. {\bf A}} \textbf{\bibinfo{volume}{60}},
  \bibinfo{pages}{1519} (\bibinfo{year}{1999}).

\bibitem{Chomaz:2000}
\bibinfo{author}{\bibfnamefont{Ph.}~\bibnamefont{Chomaz}},
\bibinfo{author}{\bibfnamefont{V.}~\bibnamefont{Duflot}},
\bibnamefont{and}
\bibinfo{author}{\bibfnamefont{F.}~\bibnamefont{Gulminelli}},
  \bibinfo{journal}{Phys. Rev. Lett.} \textbf{\bibinfo{volume}{85}},
  \bibinfo{pages}{3587} (\bibinfo{year}{2000}).


\bibitem{Bugaev:2000}
\bibinfo{author}{\bibfnamefont{K.~A.}~\bibnamefont{Bugaev}},
\bibinfo{author}{\bibfnamefont{M.~I.}~\bibnamefont{Gorenstein}},
\bibinfo{author}{\bibfnamefont{I.~N.}~\bibnamefont{Mishustin}},
\bibnamefont{and}
\bibinfo{author}{\bibfnamefont{W.}~\bibnamefont{Greiner}},
  \bibinfo{journal}{arXiV:nucl-th/0007062}  (\bibinfo{year}{2000}).

\end{thebibliography}
\end{document}